# A Computational Analysis of the Function of Three Inhibitory Cell Types in Contextual Visual Processing

Jung H. Lee[1*], Christof Koch[1] and Stefan Mihalas[1*]

[1] Allen Institute for Brain Science, Seattle, WA, USA

*Corresponding authors

E-mail: jungl@alleninstitute.org and stefanm@alleninstitute.org

## Abstract

Most cortical inhibitory cell types exclusively express one of three genes, parvalbumin, somatostatin and 5HT3a. The visual responses of cortical neurons are affected not only by local cues, but also by visual context. As the inhibitory neuron types have distinctive synaptic sources and targets over different spatial extents and from different areas, we conjecture that they possess distinct roles in contextual processing. We use modeling to relate structural information to function in primary visual cortex (V1) of the mouse, and investigate their role in contextual visual processing. Our findings are threefold. First, the inhibition mediated by parvalbumin positive (PV) cells mediates local processing and could underlie their role in boundary detection. Second, the inhibition mediated by somatostatin-positive (SST) cells facilitates longer range spatial competition among receptive fields. Third, non-specific top-down modulation to interneurons expressing vasoactive intestinal polypeptide (VIP), a subclass of 5HT3a neurons, can selectively enhance V1 responses.

## Introduction

Inhibitory cells have been considered crucial in regulating neural activity, but a mechanistic understanding of their functional roles remains elusive. In part this is because the inhibitory cell types are multifarious in their morphologies and characteristics. Recent developments of transgenic and optogenetic manipulation demonstrates that the diversity of interneurons can be mapped onto a finite number of classes (1–3). For instance, Rudy et al. (4) found that nearly all neocortical inhibitory cell types express one of the three genes PV, SST and 5HT3a exclusively, with roughly 40% of 5HT3a cells expressing VIP. Moreover, PV, SST and VIP have distinctive connectivity (5). PV cells inhibit pyramidal cells and themselves, SST cells inhibit all other cell types except themselves, and VIP cells exclusively suppress SST cells (Fig. 1A).

Recent experiments have corroborated a link between these three inhibitory cell types and distinctive functions. PV cells regulate sensory signal processing in the barrel cortex (6) and modulate the gain of visual neurons (7). SST cells participate in the surround suppression (8). VIP cells thought to be associated with disinhibition of pyramidal cells (5) are activated during negative feedback (9) and mediate top-down modulation to V1 (10).

We here use a computational model of V1 to investigate how the three inhibitory cell types modulate cellular responses in contextual visual signal processing. We focus on studying their roles in regulating interactions among visual neurons with distinctive receptive fields (RFs), for a better knowledge of the interactions among RFs can expose the neural mechanisms underlying contextual spatial processing (11). To examine the role of cell type specific connectivity among inhibitory neurons, we used a minimalistic approach in which we started from an existing columnar model (Potjans and Diesmann 2014, Wagatsuma et al 2013), and modified only a small number of variables by adding superficial layer circuits incorporating PV, SST and VIP cells. We modeled multiple such nearby columns by assuming that each column responds to a unique RF.

Our simulations demonstrate individual roles for each of the three inhibitory cell types in processing spatial scene context. Firstly, PV cells control the gain of V1 responses and shape the spatial profile of the model response, and could account for the insensitivity of V1 neurons to homogeneous surfaces (11). Secondly, SST cells facilitate the competition between objects in the visual scene, thereby effectively enhancing figure-ground contrast. Lastly, a non-specific activation of VIP cells can selectively enhance the responses to preferred stimulus due to coordination among the three inhibitory cell types.

## Results

### Input/Output relations for a microcircuit with the three inhibitory cell types

We here consider the superficial layer circuits consisting of pyramidal (Pyr), PV, SST and a subset of 5HT3a positive neurons (about 40%) that express VIP (4). We adopt the circuit diagram reported by Pfeffer et al. (5) for the superficial layer 2/3 (Fig. 1A).

We first perform a qualitative analysis to better understand the interactions among these four cell types in isolation, using a reduced set of four firing rate equations (see S1 Text). The red, black and blue lines in S1 Fig. represent the stable, unstable steady and periodic solutions of the firing rate equation for the population of excitatory cells. They indicate distinctive effects of SST and VIP cell activity on layer 2/3 Pyr cell activity. As expected, SST cells suppress Pyr cell activity (S1A Fig.). The steep decrease of Pyr cell activity can be attributed to the rapid increases of SST cell activity (S1B Fig.). When the input to SST cells is higher than ~365 $pA$, SST cell activity is strong enough to silence all other cell types. After this point, SST cells receive no internal interactions but are purely driven by external inputs: the firing rate can be predicted by its neuronal gain function (see S1 Text). Pyr cell activity reduces more slowly (S1C Fig.), when we reduce SST cell activity by decreasing the coefficient of the gain function (see S1 Text), confirming that the steep decrease of Pyr cell activity results from the rapid increases of SST cell activity.

Contrariwise, increasing the input to VIP cells disinhibits Pyr cells (S1D Fig.). As the input to VIP cells increases, SST cell activity decreases (S1E Fig.). As a result, Pyr cells receive net reduced inhibition since VIP cells only weakly inhibit Pyr cells. On the other hand, if the external input to VIP is below the threshold (360 $p$A), there is no inhibition impinging onto SST cells, making SST cell activity be strong enough to suppress Pyr cell activity. We also note that SST cells become quiescent at the critical point when Pyr cell activity starts decreasing (S1D Fig.). That is, if VIP cell activity continues to grow after this point, its effect on Pyr cells becomes purely inhibitory. Without inhibition from VIP to SST cells, the input to VIP cells suppresses Pyr cells (S1F Fig.), confirming that VIP cells can disinhibit Pyr cells by suppressing SST cells. To better understand how these cell types contribute to contextual information processing, we use computational models of V1 and discuss simulation results below.

**Dynamic interaction among superficial layer cells can be critical for the responses of the whole column**

We first embed the superficial circuit into the columnar model of Wagatsuma, Potjans, Diesmann, Sakai and Fukai (12), consisting of 19,294 leaky integrate-and-fire (LIF) units stimulated by barrages of spikes generated with a Poisson process (see Methods for details). We also use their inter-laminar connectivity scheme (Fig. 1B), with the added assumption that all three superficial interneurons are treated the same with regard to inter-laminar connections (see Methods).

We examine the responses of this refined column to transient thalamic inputs onto layer 4 and layer 6 (Potjan and Diesman 2014), averaging cell type activity over 100 independent simulations (Fig. 2A). Two observations are germane here. First, excitatory (referred from now on as E) layer 4 cells relay thalamic excitation to other layers. After L4 activation, signals propagate to 2/3 and 5, which is largely consistent with feedforward activation (13,14). The onset of inhibitory (referred as I) layer 6 cell activity occurs almost simultaneously with layer 5 E cell activation, resulting from the excitation from layer 2/3 Pyr cells. Second, in the superficial layer, all cell activity is enhanced by thalamic inputs, but the characteristic response times are cell type specific. The onset of VIP cells is the earliest and is followed by Pyr, PV and SST cell activation. This is not surprising since VIP cells receive weak inhibition unless SST cells are active according to the connectivity (Fig. 1A). Such early activation of VIP cells justifies the delayed activation of SST cells because VIP cells inhibit SST cells.

Our qualitative analysis suggests that activation of SST can suppress Pyr cell activity. Indeed, the increased inputs to SST cells does reduce Pyr cell activity in the superficial layer (Fig. 2B). More importantly, all other layer activity is also reduced, suggesting that layer 2/3 Pyr cells can drive all other layer cells. Similar results occur when inputs to VIP cells are reduced (Fig. 2C). Figure 2D shows that Pyr cell activity increases by stimulating VIP cells to fire more strongly, consistent with our qualitative analysis (S1 Fig.).

**Model column responses are contextual and dependent on intercolumnar connections**

We next consider a multi columnar model of V1, combining 13 (instead of 8 as in (12)) of these columns into an one-dimensional arrangement (Fig. 3A).

Since surround suppression, the best studied inter-receptive field interactions within V1, is mediated by long-distance horizontal connections among superficial layers (8), we analyze how the three layer 2/3 inhibitory cell types contribute to intra- and inter-columnar interactions. We assume that each cortical column is associated with an individual receptive field (RF) and that all columns are connected with one another through superficial-superficial connections only (Fig. 1C). We implement two types of di-synaptic inhibitory connections (Fig. 1C; di-synaptic because excitation terminates onto interneurons that, in turn, inhibit their postsynaptic local targets): one is a long-range excitatory connection targeting SST cells, and the other a short-range connection targeting PV cells (8). We also include short-range excitatory Pyr-Pyr and inhibitory PV-Pyr connections among nearest neighbor columns (Fig. 1C). In the following, we keep all intra-columnar connections fixed, varying the number of inter-columnar connections and external input strengths to layer 2/3 cell types (see Methods).

Boundary detection is of fundamental importance to visual perception. Most boundary detection schemes identify discontinuities in the image, which can range from discontinuities in luminosity (edges) to discontinuities in higher order statistics (texture boundaries). V1 neurons are generally insensitive to homogenous surfaces (11), which can be explained by inhibition from nearby cells with similar responses. To model which inhibitory cell type predominantly contribute to this process, we study how our 13 columns model responds to a simple figure-ground stimulus. Columns 5-9, corresponding to the "figure", are considered edge- and surface-columns, with the corresponding four neighbor on each side being "ground" (Fig. 3A). The corresponding thalamic cells fire at 80 and 40 Hz, respectively, 400-500 msec after onset of simulations. As expected, layer 2/3 pyramids respond prominently within columns corresponding to the "figure" (S2 Fig.), and only weakly in "ground" columns. Following stimulus offset, their response wanes to spontaneous firing.

To compare responses among columns, we normalize the column-specific outputs (firing rates of layer 2/3 Pyr cells) to the mean value of the two edge-column outputs in each simulation during the stimulation period (400-500 msec). The mean response and standard errors from 100 independent simulations of layer 2/3 Pyr cells are displayed in Fig. 3B. The reference value is the mean value of edge column responses marked by arrows when the Pyr-Pyr inter-columnar connection probability is 6.6%. As seen in the figure, the figure-responses are context-dependent, and the exact spatial profile is determined by the intercolumnar interactions. When Pyr-Pyr interaction is strong (red lines), the response to the surface is stronger than that to the edges, whereas edge columns generate stronger outputs (blue and black lines in Fig. 3B) when Pyr-Pyr interaction is feeble.

However, this comparison only shows the average level of outputs over 100 trials, and does not necessarily suggest the reliability of the contextual responses generated in each trial. Thus, we normalize the model outputs to the mean responses to the edges in each simulation and display them in Fig. 3C to confirm that inter-columnar connections can induce contextual responses on a trial-by trial basis; we use this trial-by-trial basis normalization for the rest of Fig. 3 and S3 Fig. The inhomogeneous responses of model columns, dependent on intercolumnar connections, can be explained by the boundary effects induced by the discontinuity existing at the edges. The edge columns receive less excitation and inhibition through inter-columnar connections than surface columns do, since the background induces less Pyr cell activity in the corresponding columns. That is, edge columns will have less disynaptic inhibitory inputs than surface columns if the net intercolumnar inputs are effectively inhibitory. We test this hypothesis by increasing the connection probability for Pyr-PV and PV-Pyr, both of which enhance intra-columnar inhibition. As expected, the response of the surface columns is reduced when its strength is increased (Figs. 3D and E). In addition, the short-range inhibition can reduce surface column responses more effectively when the Pyr-Pyr connection probability is lowered to 1% (S3 Fig.). These results do confirm our hypothesis. However, it is possible that the globally enhanced inhibition from PV to Pyr cells is capable of generating edge dominant responses. To test this possibility, we perform the same simulations but with enhanced background inputs to PV cells; specifically, we increase the frequency of spike trains carried by a single external fiber to PV cells. As seen in Fig. 3F, the responses of surface columns are not reduced when the inhibition is globally enhanced, confirming that the intercolumnar inhibition mediated by PV cells is necessary for realizing edge-dominant responses.

Next, we examine if the functional long-range inhibition, mediated by Pyr cells making long-range connections onto SST cells (Fig. 1C), can modulate the spatial profile of column responses by increasing the connection probability for Pyr-SST cells. The strengthened Pyr-SST connections enhance the response of surface columns (Fig. 3G), which is strikingly different from edge-dominant responses (Fig. 3E). Specifically, column 7, which receives the strongest excitation due to inter-columnar excitation, generates twice the response of the edge columns. Once again, the responses in Fig. 3G are all relative to the edge column responses in each simulation condition. In fact, the spiking activity of Pyr cells in background columns becomes less as the connections from Pyr to SST cells increases (Fig. 3H); this indicates that Pyr-SST cell connections can effectively reduce surround suppression to edge- and surface-columns.

This raises the question of why does long-range inhibition via SST cells behave differently from short-range inhibition mediated by PV cells? All SST cells in the figure-columns receive inter-columnar excitation from 4 figure and 4 ground columns. That is, there is no spatial gradient of inhibition among figure-columns, suggesting that long-range inhibition is insensitive to the boundary effects inducing edge-dominant responses. In addition, once inter-columnar connections from Pyr to SST cells are increased, PV cell activity decreases due to the inhibition from SST to PV cells. Together, short and long-range inhibition generate paradoxical effects.

**The coordination between SST and VIP cells can selectively suppress neural responses to the backgrounds**

Figure 3G indicates that SST cells can effectively control the competition over large spatial scales between figure and ground. However, a strong competition may be undesirable under certain circumstances. For instance, when two objects are in close proximity, so that the two corresponding columns could inhibit each other via Pyr-SST cell connections, the dominant object could prevent the non-dominant one from evoking a response. Therefore, we examine how long-range inhibition mediated by SST cells modulate our model's responses to multiple objects, here a dominant and a non-dominant object embedded in the background (Fig. 4A). The former induces 80 Hz firing in its associated thalamic cells and targets cortical column 5, while the latter triggers 60 Hz thalamic firing, projecting to column 8. All other columns receive thalamic afferents firing at 40 Hz, corresponding to ground. In the following, we designate a column as dominant, non-dominant and ground columns according to their thalamic sources below.

We evaluate the effects of SST-cell mediating long-range inhibition on Pyr cell activity in dominant and non-dominant columns by increasing the connection probability for Pyr-SST cells by a factor of 20. After performing 100 independent simulations with fixing connection probability for Pyr-SST to be 0.2%, we calculate the mean firing rate of Pyr cells for ground and use it as a reference value. The normalized responses to the two stimuli do not change noticeably until Pyr-SST is increased by ten, from 0.2 to 2% (Fig. 4B). Increasing inhibition further accentuates the response of the dominant stimulus. This result can be explained by the reduction of surround suppression from background/non-dominant to the dominant columns. We also note that Pyr cell activity in the non-dominant column also increases if it is normalized to the responses to ground (the dashed black line in Fig. 4B). This is somewhat surprising since the spiking activity of Pyr cells in the dominant column, which can project di-synaptic inhibition to Pyr cells in the non-dominant column, is greatly enhanced.

To investigate the mechanisms underlying the minimal reduction of non-dominant column responses, we calculate the firing rates of SST cells in the dominant, non-dominant and ground columns. We expect disparate SST cell activities among columns, since SST cells in the columns receive different synaptic inputs. Figure 4C shows the results depending on the connection probability for Pyr-SST cells. All firing rates are normalized to the background-evoked SST cell activity, with the lowest connection probability (2%) for Pyr-SST cells. The disparity in SST cell activity among columns is not high, suggesting that the long-range inhibition suppresses Pyr cells equally across all columns. Such uniform inhibition can most effectively reduce Pyr cell in the ground columns and reduce surround suppression to both dominant and non-dominant columns; this reduced surround suppression to the non-dominant column may compensate for the enhanced di-synaptic inhibition from the dominant column.

It should be noted that VIP cell activity is the highest in the dominant column and the lowest in the ground columns (Fig. 4D). In other words, VIP cells can provide stronger inhibition to Pyr cells in the non-dominant column than in the ground columns, which renders SST cell activity relatively uniform across all columns (Fig. 4C).

## VIP cells can work as a gate keeper for the contextual information from higher order cortical areas

Finally, we wish to elucidate the potential mechanisms underlying an elegant experiment that genetically targeted subset of neurons in cingulate cortex in the mouse (Cg), which directly project to V1, activating or inactivating these via optogenetic tools (10). This enhanced or reduced orientation-tuning of V1 neurons, possibly via local activation of superficial VIP neurons. Remarkably, Cg activation also enhanced behavioral performance of the mice in an orientation discrimination task (10). The author noted that it likely that the laser light stimulating the ChR2 expressing neurons activated the entire Cg, leading to the possibility that the observed gain modulation is induced by non-specific top-down signaling. How could non-specific VIP cell activation selectively enhance neural responses to the preferred orientation?

To gain insight into the mechanisms of top-down gain modulation via VIP cells, we consider V1 responses to an object occupying the RF of a single column. As mouse V1 lacks orientation columns, these simulations do not directly explain the sharpened orientation tuning curve [10]. However, we aim to better understand potential mechanisms by which non-specific VIP cell activation selectively enhances V1 responses. In addition, mouse V1 may have 'effective' orientation columns [15], suggesting that our simulation results with distinct RFs can be a good indicator for gain modulation of orientation tuning curve. For this simulation, we assume that column 7 is preferentially excited (here by lateral geniculate nucleus cells firing at 80 Hz), while all other columns only receive 40 Hz geniculate input (Fig. 5A). In addition, cell-type specific external inputs are homogeneous in all columns; this ensures that our model simulates the non-specific top-down signaling. We observe (Fig. 5B-F) that layer 2/3 Pyr cell activity in the preferred column is strongly related to the amplitude of synaptic inputs to VIP cells.

Furthermore, the responses in non-preferred columns are insensitive to those inputs. We also observe that the effects of non-specific activation of VIP cells are dependent on overall SST cell activity. The gain becomes more pronounced as the inputs to SST cells becomes stronger. These results are consistent with the qualitative analysis (S1E Fig.), suggesting that VIP cells disinhibit Pyr cells by suppressing SST cell activity. If SST cell activity is too low, the activation of VIP cells cannot enhance Pyr cell activity but decreases it. It should be noted that Cg indeed activates SST cells as well (10), such that the simulations in Figs. 5D and E are probably closer to the experimental conditions.

## Discussion

In this study we propose a refined multiple column model of visual cortex which incorporates three inhibitory cell types and cell-type specific connectivity among them. We use these newly developed models to elucidate the functional roles of the three inhibitory cell types in contextual visual signal processing in V1, and our computational analysis indicates cell-type specific functions.

### Inter-columnar interactions play a role in processing contextual information included in visual scene

Depending on the inter-columnar connections, either edge-responding or surface-responding columns can generate dominant responses despite the equivalent level of thalamic inputs (Fig. 3). More importantly, PV and SST cells in our models participate in processing spatial contexts in the visual scenes, but their functional roles are distinctive. Short-range inhibition mediated by PV cells generates local spatial gradient of inhibition, allowing V1 neurons to respond to discontinuities induced by the edges of the objects. In contrast, long-range inhibition mediated by SST cells reduces the responses to ground or non-preferred stimuli but increases the responses to preferred stimulus. This selective enhancement, similar to a "winner-take-all" operation, can be explained by the reduction of surround suppression impinging onto V1 neurons responding to the preferred stimulus (Fig. 3H). We also note that VIP cells can regulate the amount of surround suppression projecting to specific V1 neurons depending on the thalamic inputs (Fig. 4). Edge-dominant responses, induced when inter-columnar inhibition is effectively inhibitory, can be useful in detecting the edges of visual objects. This behavior indeed is consistent with the insensitivity of V1 neurons to extended surfaces (11). The effective winner-take-all operation introduced by long-range inhibition helps visual neurons distinguish an object from others or from background.

**The three inhibitory cell types coordinate to allow top-down modulation**

Our model V1 responses are modulated by the external inputs to VIP cells in a manner consistent with the hypothesis that VIP cells mediate top-down signaling to V1 (Fig. 5). Three observations of our simulation results are noteworthy. First, the spiking activity of layer 2/3 Pyr cells responding to the preferred stimulus is selectively enhanced by non-selective activation of VIP cells. That is, even when top-down signaling affects a large portion of V1, the induced effects are pronounced only in the target area, suggesting that top-down signaling need not be strictly target-specific. The same stimulus specificity without corresponding spatial specificity of the afferent input also applies to neuromodulatory input such as acetylcholine. We also note that the lateral inhibition mediated by PV and SST cells can control responses in columns responding to non-preferred stimuli (S4A and B Figs.).

Second, the gain control mediated by VIP cells is dependent on the overall SST cell activity (Fig. 5). The more active SST cells are, the higher the effective gain. Interestingly, both feeble and prominent effects induced by the activation of VIP cells were reported. Lee et al. (7) found mild changes in the firing rates when VIP cells were directly activated. In contrast, Zhang et al. (10) found a multiplicative gain modulation when Cg, which innervates VIP cells in V1, was activated. The dependency of the gain on SST cell activity in our model suggests that the difference between the two experiments could be attributed to the fact that SST cell also receives EPSCs from Cg. Thus, Cg activation corresponds to the simulation results with the higher inputs to SST cells. In contrast, when animals passively watch visual stimuli as in Lee et al. (7), Cg may not be active, and thus SST cell activity could be too low for the disinhibitory control induced by VIP cells.

Third, the enhanced external inputs to PV cells globally reduce Pyr cell activity (S4C Fig.). This is consistent with the subtractive effects experimentally observed (7), which can sharpen the tuning curve; see (15) for recent discussion. Again, it should be noted that Cg activation induces EPSCs in PV cells as well, suggesting that Cg may activate the PV cells to suppress the responses of Pyr cells responding to the non-preferred stimuli.

In brief, our model proposes that Cg excites all three inhibitory cell types to promote the coordination among the three cell types, allowing non-specific top-down signaling to have target-specificity.

**Comparison to other models**

Although inhibitory cell types are very diverse, only a few models considered two inhibitory cell types (FS and non-FS interneurons). Hayut et al. (16) studied interactions among Pyr, FS and low-threshold spiking (LTS) cells using firing rate equation. These two inhibitory cell types were also incorporated into the single column consisting of biophysically detailed neurons to study the underlying mechanisms of cortical rhythms (17), and a more recent modeling study (18) suggested that LTS cells are associated with deep layer beta rhythms, inspiring more abstract models focusing on the two inhibitory cell types' contribution to interlaminar interactions (19,20).

Our model considers three major inhibitory cell types and how they contribute to inter-columnar interactions. It ignores many known complications. Examples of these include cell type-dependent mechanisms such as firing rate adaptation (21), multiple inhibitory cell types in other layers (22), dynamic synapses with short- and long-term plasticity (23) and more sophisticated thalamic input (24). However, we feel that at this point in the exploration of cortex, it is best to proceed step-by-step in a systematic manner rather than generating ill-understood biophysical models with very large degrees of freedom. The impact of all such simplifications will be addressed in future studies. Currently, our institute is collecting necessary data to incorporate them into the next generation computational models of V1, and we will probe the updated models to study neural correlates underlying contextual visual information. Also we plan to incorporate cell-type specific cellular mechanisms to further study the functional roles of inhibitory cell types by using generalized LIF neuron model capable of reproducing *in vitro* physiological data.

## Methods

Our model is based on the multiple column model proposed by Wagatsuma et al. (12). In the original model, the eight columns interact with one another via excitatory synaptic connections between superficial layers. Those intercolumnar connections target excitatory and inhibitory cells. Excitatory-excitatory connections reach the nearest columns only, whereas excitatory-inhibitory connections reach all other columns. Here we modified this original model by incorporating the three inhibitory cell types in superficial layers and their cell-type specific connectivity within and across columns to study functional roles of each type in interactions across columns.

We used the peer-reviewed simulation platform "NEST" (Gewaltig & Diesmann, 2007) to build a refined model. All cells in our model are identical "leaky-integrate-and-fire" (LIF) neurons whose postsynaptic currents decay exponentially, and we used NEST-native neuron models. Specifically, we modeled superficial layer cells and other layer cells using "iaf_psc_exp_multisynapse" and "iaf_psc_exp" neuron models, respectively. These two neuron models are identical in terms of internal dynamics for integration and spiking, but the former allows multiple synaptic ports, each of which can have distinctive postsynaptic dynamics. The multiple postsynaptic dynamics are necessary for neuron models to integrate synaptic inputs from multiple types of presynaptic sources. Table 1 shows the parameters for neurons and synapses used in our model.

Each cell type in superficial layers of our model has specific presynaptic sources and postsynaptic targets, as reported in Pfeffer et al. (2013). To incorporate the reported cell-type specific connectivity, we estimated cell-type specific population size, connectivity and postsynaptic dynamics in superficial layers. All other layers are equivalent to those in Wagatsuma et al. (2013) with one exception; see Table 1. Below we illustrate the details about our estimates in superficial layers.

**Population size**

We split superficial layer inhibitory cells into three populations according to Rudy et al. (4). First, we set 24% of inhibitory cells in an individual column in Wagatsuma et al. (12) to be vasoactive intestinal peptide-positive (VIP) cells, since Rudy et al. (4) suggested that 60% interneurons in superficial layers express 5HT3a and 40% of them also express VIP. Second, we set 30% cells to be somatostatin (SST) cells using the average faction found in all layers. Third, we set the rest to be parvalbumin (PV) cells, which are the most common inhibitory cells in the cortex.

Pfeffer et al. (5) estimated that PV, SST, VIP and unknown types constitute 36%, 30%, 17% and 16% of inhibitory neurons, respectively. If we include 16% unknown types in the PV cell group, our estimates are consistent with their findings. Especially, the difference in VIP cell fraction can be attributed to the fact that Pfeffer et al. (5) measured these values in layer 5 as well, in which VIP cells are less common than those in layer 2/3; thus, their estimates of VIP cell fraction will be lower than those in layer 2/3 alone.

**Connectivity among cell types in superficial layers**

Wagatsuma et al. (12) connected excitatory and inhibitory cells in superficial layers by specifying 4 connection probabilities $P_{EE}$, $P_{EI}$, $P_{IE}$ and $P_{II}$ We used $P_{EE}$ and $P_{EI}$ for recurrent connections among pyramidal cells and excitatory projections impinging onto the three inhibitory cell types. That is, the three inhibitory cell populations are equally connected to the pyramidal cell population; the number of synaptic connections between populations is dependent on the size of postsynaptic cell population, as suggested in Potjans and Diesmann (25).

We connected the three inhibitory cell types to pyramidal cells using the cell-type specific individual neuronal contribution (INC) on inhibition in mouse visual cortex (5). Specifically, we first computed the total number of synapses from inhibitory cells to pyramidal cells used in Wagatsuma et al. (12). Then we split them into three populations using the connection probabilities reported in Pfeffer et al. (5) as weighting factors; see Table 1. In the same way, we implemented recurrent connections among the three inhibitory cells. Figure 1 illustrates the cell-type specific inhibitory connections used in our model. We note that this connectivity is adopted from the circuit diagram proposed by Pfeffer et al. (5). Here we added two more connections VIP-Pyr and PV-VIP into them. VIP-Pyr connection was added to capture the functional roles of inhibition projected from VIP cells to pyramidal cells, and PV-VIP cell connection was added due to the reported high INC value, as detailed in the next subsection.

**Postsynaptic currents in superficial layers**

We also approximated cell-type specific postsynaptic currents by estimating peak currents and decay time constants from data reported in Pfeffer et al. (5). These two factors are sufficient to determine the exact shape of postsynaptic currents in LIF neurons if postsynaptic currents decay exponentially. To estimate them for seven cell-pairs (PV-Pyr, SST-Pyr, VIP-Pyr, PV-PV, SST-PV, SST-VIP and VIP-SST), we measured the heights of row traces of inhibitory post synaptic charges (IPSQ) given in Pfeffer et al. (5) and converted them into peak currents using the reference bar. Table 1 displays our estimates. Once we obtained the peak currents, we calculated the decay time constants by dividing IPSQ values with them due to the property of the exponential-decay curve.

For SST-VIP and PV-VIP connections, the given information is not sufficient to specify both peak currents and decay time constants. We estimated the peak currents using the individual neuronal contribution (INC). Specifically, we compared INC values between SST-VIP and SST-PV pairs and between PV-VIP and PV-PV pairs. According to the definition in Pfeffer et al. (5), INC values are the products of IPSQs and connection probability $P_{con}$. As a result, one condition exists: the three unknown variables given below:

$$\frac{INC_{SST-VIP,PV-VIP}}{INC_{SST-PV,PV-PV}} = \frac{P_{con} \times IPSQ_{SST-VIP,PV-VIP}}{P_{con} \times IPSQ_{SST-PV,PV-PV}} = \frac{P_{con} \times Peak_{SST-VIP,PV-VIP} \times \tau_{SST-VIP,PV-VIP}}{P_{con} \times Peak_{SST-PV,PV-PV} \times \tau_{SST-PV,PV-PV}}.$$

We first assumed that the two pairs of connections originating from the same presynaptic cells have the same decay time constant to remove one unknown variable. Then, one constraint should be satisfied by the multiplication of connection probabilities $P_{con}$ and peak currents. Since we used connection probabilities reported in Pfeffer et al. (5) as weight factors to distribute synaptic connections, the impacts of connection probabilities should be minimal. Thus, we chose $P_{con}=1$ for both SST-VIP and PV-VIP connections and adjusted peak currents properly, which are also given in Table 1. We noted that those values are roughly consistent with the ratio of raw IPSQ peaks provided in Pfeffer et al. (5).

The estimated decay times (see Table 1) are much longer than 0.5 msec used in Wagatsuam et al. (12) and also Potjans and Diesmann (25). Consequently, the pyramidal cells in superficial layers receive enhanced inhibition. To compensate this, we also lengthened the decay time of excitatory connection in superficial layers to 2 msec (16) by keeping the peak currents of excitatory connections at the same level as in Wagatsuma et al. (12).

**Interlaminar and intercolumnar connections**

Superficial layer cells can also interact with other layer cells. For such interlaminar interactions, we ignored individual types of inhibitory cells in superficial layers. That is, all three inhibitory cells types in superficial layer cells are treated equally by other layer cells, and they equally project inhibition to all other layer targets; we used the same connection probabilities specified in Wagatsuma et al. (12) to connect cell populations across laminar layers.

Although synaptic connections across columns are poorly understood, a line of studies (8,26–28) suggests that superficial layers are connected with one another, via intercolumnar connections, which is consistent with the model proposed by Wagatsuma et al. (12). Thus, we implemented intercolumnar connections between superficial layers only. As in Wagatsuma et al. (12), the targets of intercolumnar connections are pyramidal cells in the nearest neighbors and inhibitory cells in both neighbors and distant columns. Throughout this study, periodic boundary condition was used for intercolumnar connections to ensure that all columns receive equivalent synaptic inputs except selective thalamic signals. All intercolumnar connections are identical to the synaptic connections among superficial layer cells within a column except the conduction delay. Since intercolumnar connections are longer than intracolumnar connections, we introduced 5 times longer conduction delays to intercolumnar connections (Table 1).

**External background inputs and thalamic inputs**

All cells in our model receive cell type specific background inputs. Following the protocol used in Potjans and Diesmann, each cell type receives them via a fixed number of external fibers, each of which carries independent Poisson spike trains; see Table 2 for exact parameters used in our simulations. Each thalamic cell in our model projects independent Poisson spike train at the fixed rate $P_r$ to its targets in layer 4 and 6, randomly chosen according to the connection probabilities adopted from Wagatsuma et al. (2013). As in Potjans and Diesmann (2013), we connected 902 thalamic cells to a single column. The three thalamic populations were used, and all thalamic cells start firing 400 msec after the onset of simulations, and their firing lasts 100 msec unless stated otherwise.

**Acknowledgments**

We wish to thank the Allen Institute founders, Paul G. Allen and Jody Allen, for their vision, encouragement and support.

# References

1. Kepecs A, Fishell G. Interneuron cell types are fit to function. Nature. 2014 Jan;505(7483):318–26.

2. Tasic B, Menon V, Nguyen TN, Kim TK, Jarsky Ti, Yao Z, et al. Adult Mouse Cortical Cell Taxonomy by Single Cell Transcriptomics. Nat Neurosci. 2015;

3. Jiang X, Shen S, Cadwelll C, Berence P, Sinz F, Ecker AS, et al. Principles of connectivity among morphologically defined cell types in adult neocortex. Science (80- ). 2015;350(6264).

4. Rudy B, Fishell G, Lee S, Hjerling-Leffler J. Three groups of interneurons account for nearly 100% of neocortical GABAergic neurons. Dev Neurobiol. 2011 Jan 1;71(1):45–61.

5. Pfeffer CK, Xue M, He M, Huang ZJ, Scanziani M. Inhibition of inhibition in visual cortex: the logic of connections between molecularly distinct interneurons. Nat Neurosci. Nature Publishing Group; 2013 Aug;16(8):1068–76.

6. Cardin JA, Carlén M, Meletis K, Knoblich U, Zhang F, Deisseroth K, et al. Driving fast-spiking cells induces gamma rhythm and controls sensory responses. Nature. 2009 Jun 4;459(7247):663–7.

7. Lee S-H, Kwan AC, Zhang S, Phoumthipphavong V, Flannery JG, Masmanidis SC, et al. Activation of specific interneurons improves V1 feature selectivity and visual perception. Nature. Nature Publishing Group; 2012;488(7411):379–83.

8. Adesnik H, Bruns W, Taniguchi H, Huang ZJ, Scanziani M. A neural circuit for spatial summation in visual cortex. Nature. Nature Publishing Group; 2012 Oct 11;490(7419):226–31.

9. Pi H-J, Hangya B, Kvitsiani D, Sanders JI, Huang ZJ, Kepecs A. Cortical interneurons that specialize in disinhibitory control. Nature. 2013 Nov 28;503(7477):521–4.

10. Zhang S, Xu M, Kamigaki T, Hoang Do JP, Chang W-C, Jenvay S, et al. Long-range and local circuits for top-down modulation of visual cortex processing. Science (80- ). 2014 Aug 7;345(6197):660–5.

11. Albright TD, Stoner GR. Contextual influences on visual processing. Annu Rev Neurosci. 2002;25:339–79.

12. Wagatsuma N, Potjans TC, Diesmann M, Sakai K, Fukai T. Spatial and feature-based attention in a layered cortical microcircuit model. PLoS One. 2013 Jan;8(12):e80788.

13. Schroeder CE, Mehta a D, Givre SJ. 81A spatiotemporal profile of visual system activation revealed by current source density analysis in the awake macaque. Cereb Cortex. 1998;8(7):575–92.

14. Lakatos P, O'Connell MN, Barczak a., Mills a., Javitt DC, Schroeder CE. The leading sense: supramodal control of neurophysiological context by attention. Neuron. 2010;42(2):157–62.

15. Lee S-H, Kwan AC, Dan Y. Interneuron subtypes and orientation tuning. Nature. Nature 2014;508(7494):E1–2.

16. Hayut I, Fanselow EE, Connors BW, Golomb D. LTS and FS inhibitory interneurons, short-term synaptic plasticity, and cortical circuit dynamics. PLoS Comput Biol [Internet]. 2011 Oct;7(10):e1002248.

17. Traub RD, Contreras D, Cunningham MO, Murray H, LeBeau FEN, Roopun A, et al.

Single-column thalamocortical network model exhibiting gamma oscillations, sleep spindles, and epileptogenic bursts. J Neurophysiol. 2005 Apr;93(4):2194–232.

18. Roopun AK, Lebeau FEN, Ramell J, Cunningham MO, Traub RD, Whittington M a. Cholinergic neuromodulation controls directed temporal communication in neocortex in vitro. Front Neural Circuits. 2010 Jan;4(March):8.

19. Kramer M a, Roopun AK, Carracedo LM, Traub RD, Whittington M a, Kopell NJ. Rhythm generation through period concatenation in rat somatosensory cortex. PLoS Comput Biol. 2008 Jan;4(9):e1000169.

20. Lee JH, Whittington MA, Kopell NJ. Top-Down Beta Rhythms Support Selective Attention via Interlaminar Interaction: A Model. Coombes S, editor. PLoS Comput Biol. 2013 Aug 8;9(8):e1003164.

21. Kawaguchi Y, Kubota Y. GABAergic cell subtypes and their synaptic connections in rat frontal cortex. Cereb Cortex. 1997 Sep;7(6):476–86.

22. Markram H, Toledo-Rodriguez M, Wang Y, Gupta A, Silberberg G, Wu C. Interneurons of the neocortical inhibitory system. Nat Rev Neurosci. 2004 Oct;5(10):793–807.

23. Gibson JR, Beierlein M, Connors BW. Two networks of electrically coupled inhibitory neurons in neocortex. Nature. 1999 Nov 4;402(6757):75–9.

24. Jones EG. The thalamic matrix and thalamocortical synchrony. Trends Neurosci. 2001 Oct;24(10):595–601.

25. Potjans TC, Diesmann M. The cell-type specific cortical microcircuit: relating structure and activity in a full-scale spiking network model. Cereb Cortex. 2014 Mar;24(3):785–

806.

26. Adesnik H, Scanziani M. Lateral competition for cortical space by layer-specific horizontal circuits. Nature. Nature Publishing Group; 2010 Apr;464(7292):1155–60.

27. Martin K a. C, Roth S, Rusch ES. Superficial layer pyramidal cells communicate heterogeneously between multiple functional domains of cat primary visual cortex. Nat Commun. Nature Publishing Group; 2014;5:5252.

28. Muir DR, Da Costa NM a, Girardin CC, Naaman S, Omer DB, Ruesch E, et al. Embedding of cortical representations by the superficial patch system. Cereb Cortex. 2011;21(October):2244–60.
**Legends**

**Figure 1: Structure of the model.** (**A**) Each single cortical column consists of layer 2/3, layer 4, layer 5 and layer 6. We refined only the superficial layers by incorporating the three inhibitory cell types; all other layers consist of a single excitatory and a single inhibitory cell type, as in the earlier computational model (Potjans & Diesmann, 2014; Wagatsuma et al., 2013). (**B**) The interlaminar connections according to the presynaptic sources. Thick arrows show connections whose probability is higher than 10%. (C) The four-types of inter-columnar connections among superficial neurons including short-range PV and long-range SST inhibition; no other intercolumnar connections are considered in this paper. See also Figure S1

**Figure 2: Time course of cell-type specific cortical activity following transient thalamic inputs.** The thalamic input is modeled as a single 10 msec wave of spikes starting at 400 msec. (A) Response of the refined single column with standard parameters responding to the transient thalamic inputs. Thalamic inputs are projected to both excitatory (E) and inhibitory (I) cells in layers 4 and 6. Panels (**B**)-(**D**) show the same measure but with different external inputs. The non-default network values chosen are shown at the top of panels. These detailed simulations replicate our qualitative analysis of S1 Fig. : in which SST activation (**B**) or VIP inactivation (**C**) results in reduced responses, and VIP activation (**D**) results in prolonged responses. See also Figure S2

**Figure 3: Spatial context dependent responses of layer 2/3 Pyr cells.** (**A**) Stimulus layout, with a simplified "object" (corresponding to columns 5-9) superimposed onto a "ground" (the nearby columns 1-4 and 10-13; with periodic boundary conditions. (**B**) Normalized responses of Pyr cells between 400-500 msec while varying intercolumnar Pyr-Pyr connection probabilities. Error bars represent standard errors. The reference point is the mean value of outputs of edge-columns with default connection probability 6.6% for Pyr-Pyr connections. (**C**) As in (**B**), except responses are normalized to the mean value of edge responses, on a trial-by-trial basis. Similarly, Panels (**D**) and (**E**) display the normalized responses with different connection probabilities for PV-Pyr and Pyr-PV, respectively. (**F**) Column-specific outputs with three different connection external inputs to PV cells. (**G**) Dependency of normalized responses on Pyr-SST connection strength. See also Figure S3

**Figure 4: The effects of long-range inhibition on responses**. (**A**) Stimulus layout of two, one-columnar wide objects. (**B**) Firing rates of layer 2/3 Pyr cells normalized to ground-evoked responses with the lowest connection probability (0.2%). The ratio of non-dominant to ground responses is shown in the black dash line. (**C**) and (**D**) SST and VIP cell activity in all columns, respectively. The reference values in them are the background-evoked responses with the lowest connection probability (2%). Note that the normalized firing rate of SST cells does not vary across the input.

**Figure 5: The effects of top-down inputs to VIP cell onto layer 2/3 pyramids.** A single thalamic "object" excites it corresponding cortical column. (**B**) Normalized columnar responses receiving top-down input, in a spatial homogeneous manner, to both VIP and SST cells. Note that, as in previous figures, decreasing VIP activity causes an increase in Pyr cell firing. The reference value is the response of the preferred column with 8 Hz input to VIP cells. Panels (**C**)-(**F**) show the same results but with enhanced levels of input to SST cells. See also Figure S4.

**Table 1: Parameters for the network**. The table below lists the parameters regarding the structure of the model used during simulations. The same constants are maintained for all simulations unless stated otherwise. In cell-type specific connections in superficial layers, both presynaptic and post synaptic cells are one of Pyr, PV, SST and VIP; presynaptic on the left side of the arrow, postsynaptic on the right. We also strengthened excitatory connections from layer (L) 4 E to Pyr to balance the excitation from L4 E and Pyr impinging onto Pyr, as in Potjans and Diesmann (25). For all other connections, we used the same parameters used in Wagatsuma et al. (2013).

**Table 2: External background inputs**. In our simulations, all cells receive background external inputs via external fibers, each of which carries Poisson spike trains. The cell-type specific fiber number and the frequency of Poisson inputs are given below. We adopted the peak currents from Potjans and Diesmann (Potjans & Diesmann, 2014)

# Tables
**Table 1**

| Neuron Parameters | | Decay time constants (msec) | |
|---|---|---|---|
| $\tau$ | 10 msec | Pyr→Pyr | 2.0 |
| $V_{th}$ | -50 mV | PV→Pyr | 6.0 |
| $V_{reset}$ | -65 mV | SST→Pyr | 7.5 |
| $\tau_{ref}$ | 3msec | VIP→Pyr | 6.2 |
| $C$ | 250 pF | Pyr→PV | 2.0 |
| Peak currents (pA), $w \pm \delta w$ | | PV→PV | 4.3 |
| default excitatory | $175.6 \pm 17.6$ | SST→PV | 3.4 |
| default inhibitory | $-702.4 \pm 70.2$ | Pyr→SST | 2.0 |
| PV→Pyr | $-466.7 \pm 46.7$ | VIP→SST | 10.4 |
| PV→PV | $-638.1 \pm 63.8$ | Pyr→VIP | 2.0 |
| PV→VIP | $-140.04 \pm 14.0$ | PV→VIP | 4.3 |
| SST→Pyr | $-200.0 \pm 20.0$ | SST→VIP | 3.4 |
| SST→PV | $-228.6 \pm 22.9$ | default exc | 0.5 |
| SST→VIP | $-525.8 \pm 52.6$ | default inh | 0.5 |
| VIP→Pyr | $-76.2 \pm 7.62$ | | |
| VIP→SST | $-66.7 \pm 6.7$ | | |
| L4E→Pyr | $245.84 \pm 24.6$ | | |
| Synaptic delay (msec), $d \pm \delta d$ | | Neuron # | |
| Intracolumnar exc. | $1.5 \pm 0.75$ | L2/3 | E:5171  I:1459 |
| Intracolumnar inh. | $0.75 \pm 0.375$ | L4 | E:5479  I:1370 |
| Intercolumnar exc. | $7.5 \pm 3.75$ | L5 | E:1213  I:266 |
| Intercolumnar inh. | $3.75 \pm 1.88$ | L6 | E: 3599  I: 737 |
| Connection Probabilities for intercolumnar connections | | Weighting factors for cell-type specific connections | |
| Pyr-SST | 0.002 | PV-Pyr : SST-Pyr : VIP-Pyr = 1 : 1 : 0.125 | |
| Pyr-PV | 0.009 | | |
| PV-Pyr | 0.046 | PV-PV : SST-PV : VIP-SST : SST-VIP : PV-VIP = 1 : 0.857 : 0.625 : 1 : 1 | |
| Pyr-Pyr | 0.066 | | |

**Table 2**

| Number of external fibers | |
|---|---|
| L2/3 | Pyr: 1600  Inh: 1500 |
| L4 | Exc: 2100  Inh: 1900 |
| L5 | Exc: 2000  Inh: 1900 |
| L6 | Exc: 2900  Inh: 2100 |
| Background spike rate (Hz) per external fiber | |
| L2/3 | Pyr: 8.0 PV:10.0 SST:2.0 VIP:8.0 |
| All other layers | Exc: 8.0 Inh:8.0 |
| Peak currents $w \pm \delta w$ | |
| 87.9 $\pm$ 8.8 pA | |

## Supporting information

**S1 Text. Qualitative Analysis**

**S1 Fig. Qualitative analysis of the functional roles of SST and VIP cells**. The red, black and blue lines represent the stable, unstable steady and periodic solutions of the firing rate equation for the excitatory cells. (A) SST input has an overall inhibitory influence onto Pyr activity. (B) SST cell activity grows very rapidly. (C) The speed of decrease of Pyr cell activity is dependent on the gain of SST cells. (D) Importantly, external inputs to VIP cells first lead to a counter-intuitive disinhibition of pyramids before VIP activity shuts down SST firing as well as pyramidal cell activity. (E) When SST cells are quiescent, VIP input becomes purely inhibitory; only steady stable solutions are displayed in the panel. (F) VIP input cannot disinhibit Pyr cells without VIP-SST connection.

**S2 Fig.  Time course of layer 2/3 Pyr cell activity in multiple columns.** The upper panel shows layer 2/3 Pyr cell activity in surface columns (6, 7 and 8). Pyr cell activity is averaged over 100 independent simulations. Similarly, middle and lower panels show averaged layer 2/3

Pyr cell activity in edge and ground columns, respectively. All ground responses are shown in black. The red arrows represent the onset of thalamic inputs. Default parameters are provided in Tables 1 and S1.

**S3 Fig. The effects of short-range inhibition on contextual responses.** Panels **(A)** and **(B)** show the same results as Figure 3C and D but with different connection probability for Pyr-Pyr cells. To compare the shapes of the response curve, we normalize the outputs using the edge-responses for each connection probability to make all edge responses identical.

**S4 Fig. The effects of inhibition on shaping the tuning curve. (A)** and **(B)** The tuning curve can be modulated by short- and long-range inhibition across columns. **(C)** The subtractive effects is induced by enhanced inputs to SST cells.

## Supporting Information

### S1 Text

We used Wilson-Cowan type firing rate equations to have qualitative understanding of dynamics among the four cell types (Pyr, PV, SST and VIP) in superficial layers. As in our computational model, we did not consider intrinsic properties of each cell type. Instead, all cell types have distinctive connectivity. We used the F-I curve of a LIF neuron as a gain function for all cell types. Since the F-I curve of LIF neuron is well fitted to the square root-curve ($g(x) = 5.33\sqrt{x - \theta}$), we used it as the gain function. Thus, the firing rates of the four cell types can be described by Equation 1:

$$\tau_m \frac{df_e}{dt} = -f_e + 5.33\sqrt{(I_e + S_{ee}f_e - S_{ep}f_p - S_{es}f_s - S_{ev}f_v - \theta)} H(I_e + S_{ee}f_e - S_{ep}f_p - S_{es}f_s - S_{ev}f_v - \theta)$$

$$\tau_m \frac{df_p}{dt} = -f_p + 5.33\sqrt{(I_p+S_{pe}f_e-S_{pp}f_p-S_{ps}f_s - \theta)}H(I_e+S_{ee}f_e-S_{ep}f_p-S_{es}f_s - \theta)$$

$$\tau_m \frac{df_s}{dt} = -f_s + 5.33\sqrt{(I_s+S_{se}f_e-S_{ev}S_{ev} - \theta)}H(I_s+S_{se}f_e-S_{ev}S_{ev} - \theta)$$

$$\tau_m \frac{df_v}{dt} = -f_v + 5.33\sqrt{(I_v+S_{ve}f_e-S_{vp}f_p-S_{vs}f_s - \theta)}H(I_v+S_{ve}f_e-S_{vp}f_p-S_{vs}f_s - \theta)$$

(1)

, where $I_x$, $f_x$, $S_{xy}$ and $\theta = 360.0$ are the applied current, firing rate, synaptic weight from presynaptic cell $y$ to postsynaptic cell $x$ and spiking threshold, respectively; where $H$ is the Heaviside step function; where *e, p, s , v* represent Pyr, PV, SST and VIP cells, respectively. To estimate the weight $S_{xy}$, we calculated the total synaptic currents during $\tau_m = 10$ msec using the same parameters used in computational models (see Table 1). Specifically, we set $S_{ee} = 1.98$, $S_{ep} = 5.68$, $S_{es} = 3.05$, $S_{ev} = 0.12$, $S_{pe} = 0.55$, $S_{pp} = 2.28$, $S_{ps} = 0.55$, $S_{se} = 0.55$, $S_{sv} = 0.36$, $S_{ve} = 0.55$, $S_{vp} = 0.50$, $S_{vs} = 1.48$.

These equations can be considered Wilson-Cowan equation without the correction terms referring to the neurons' inability to fire during their refractory period. We ignored the correction terms since they will be small unless the neurons' firing rates are high. We numerically solved these equations and performed continuation analysis using the open-source numerical analysis package XPPAUT (Ermentrout, 2007).

## S1 Text References

Ermentrout, B. (2007). XPPAUT. *Scholarpedia*, 2(1), 1399.

# Figures
**Figure 1**

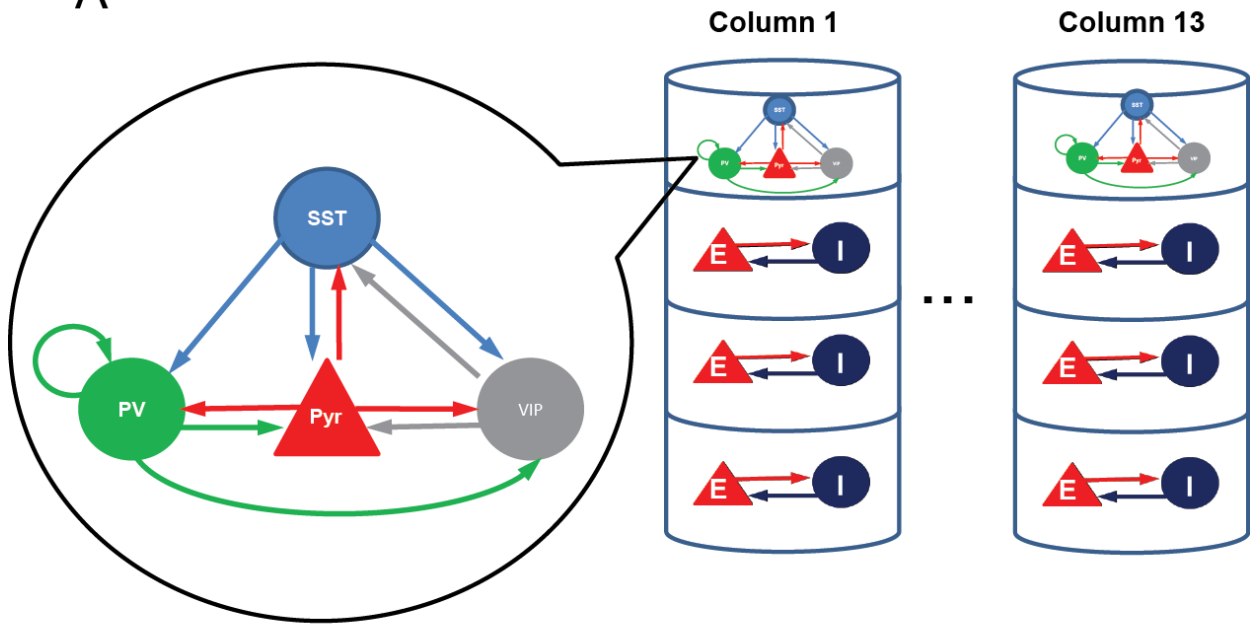

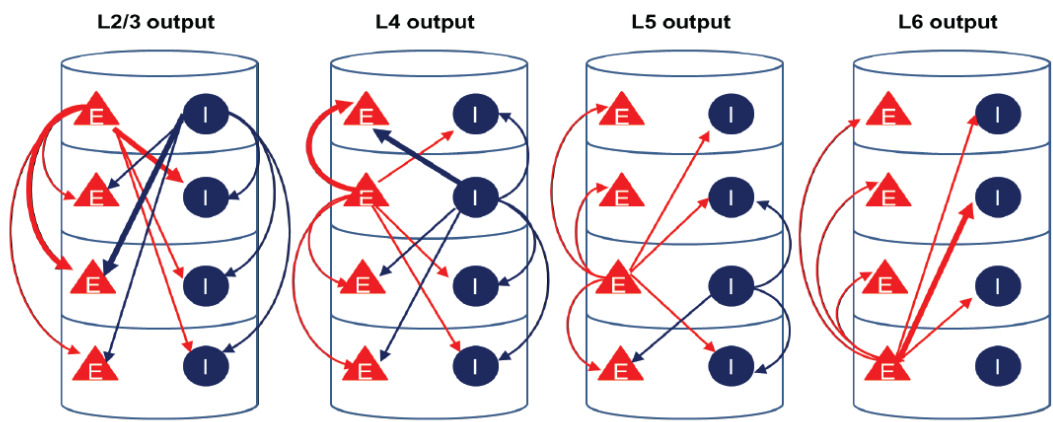

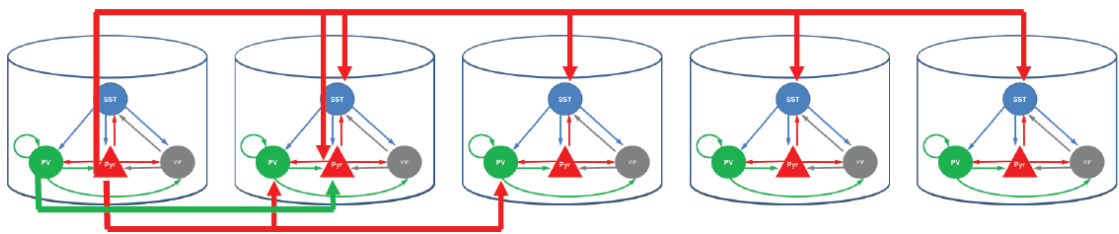

**Figure 2**

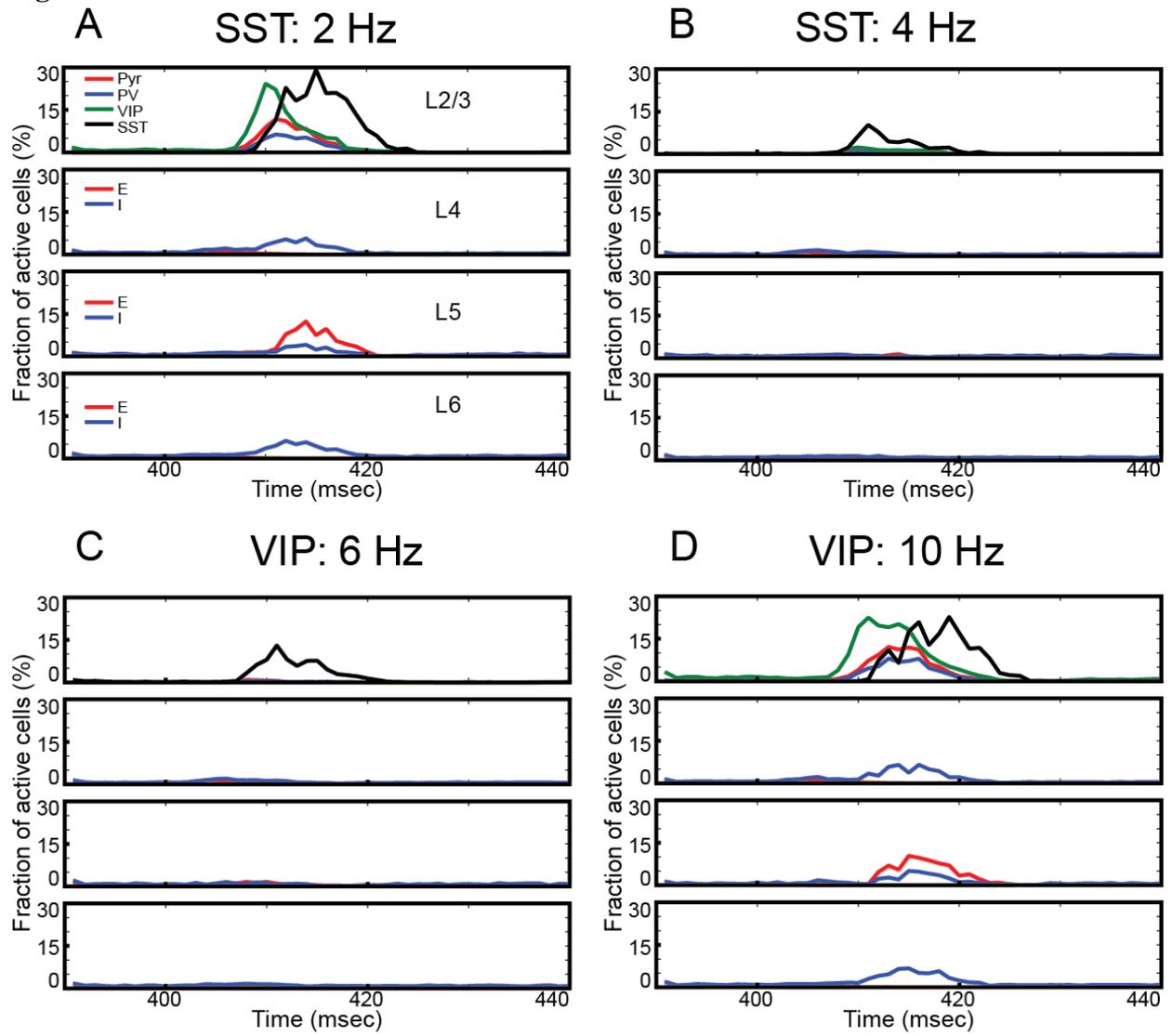

**Figure 3**

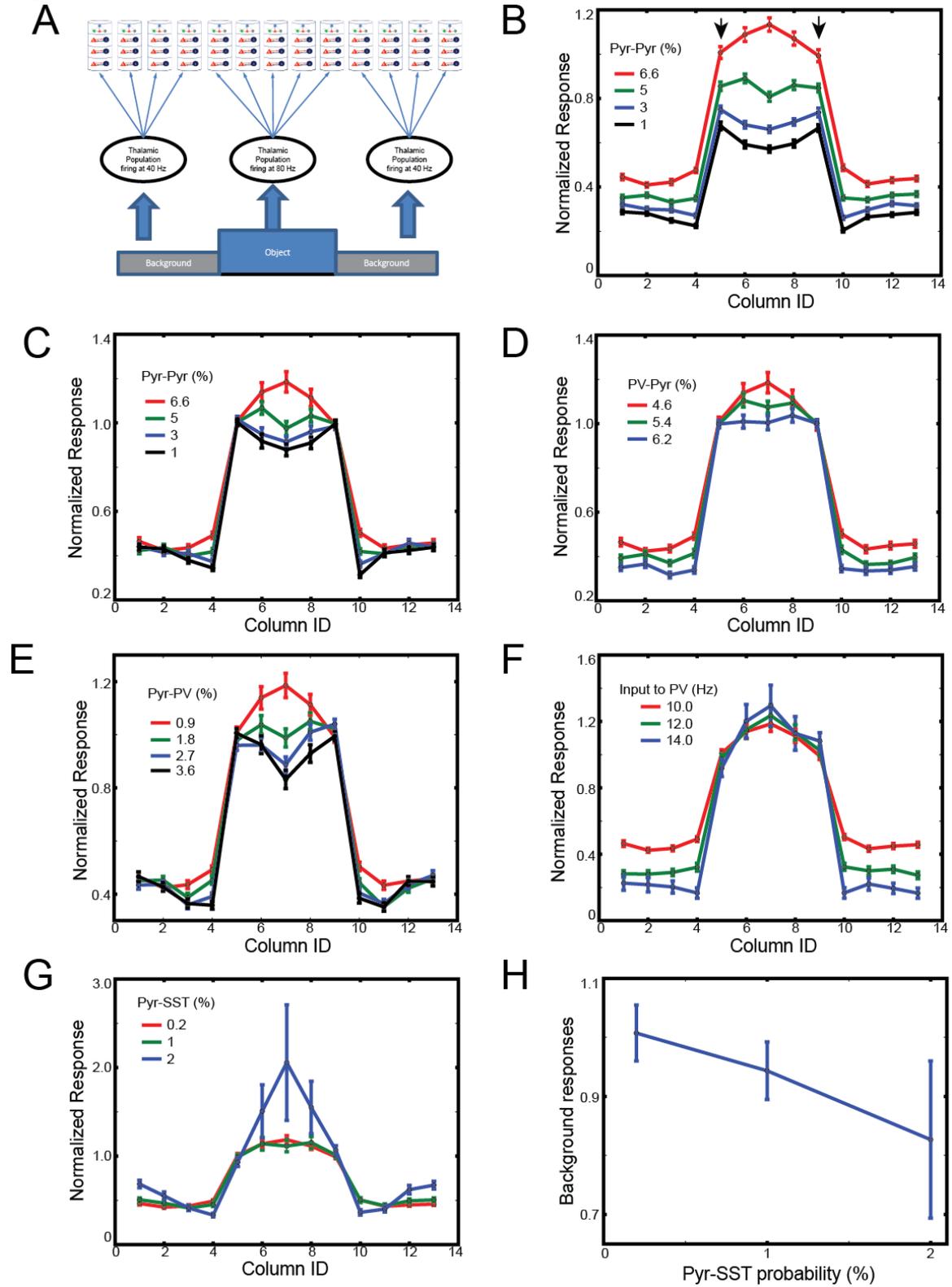

**Figure 4**

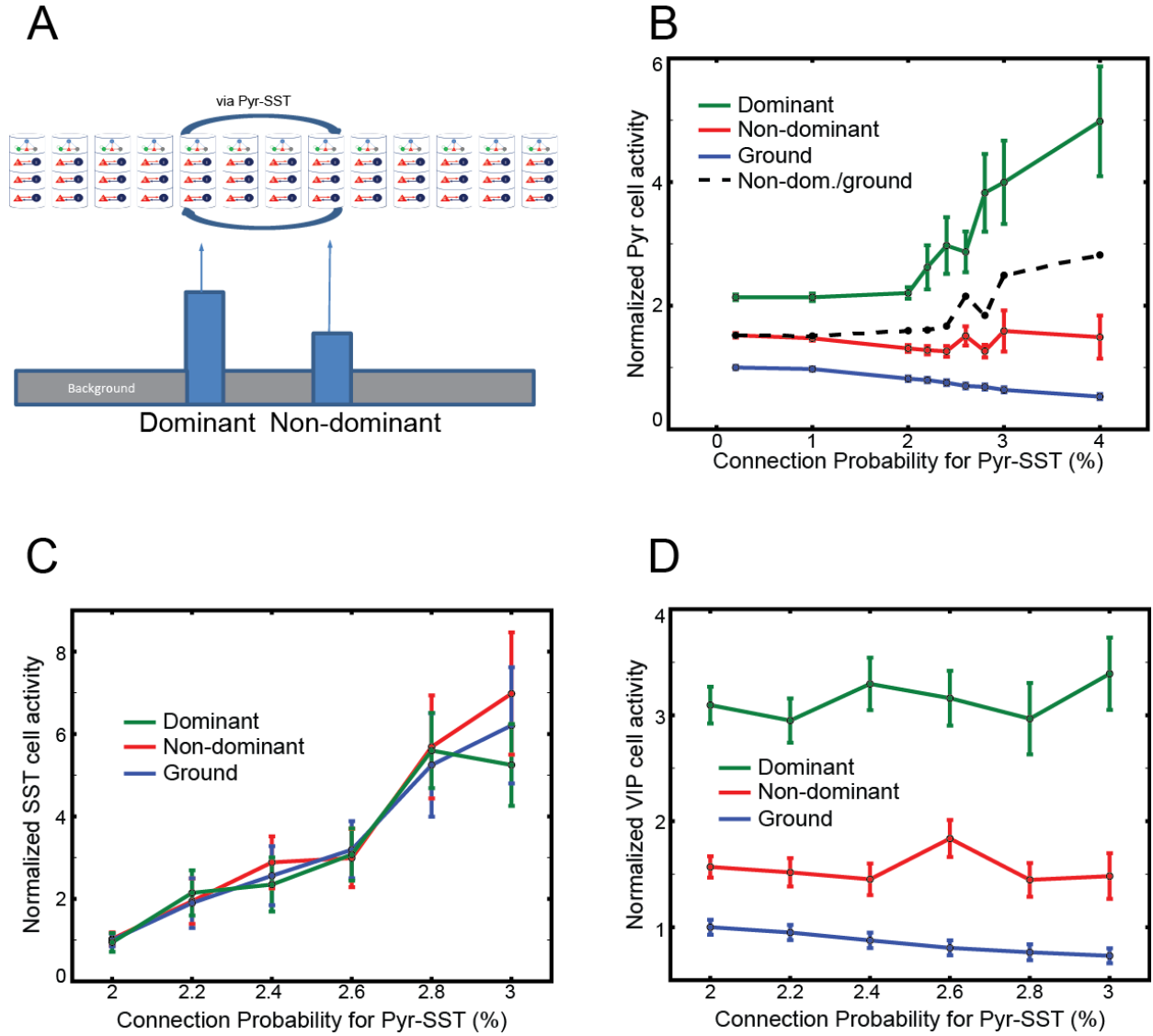

**Figure 5**

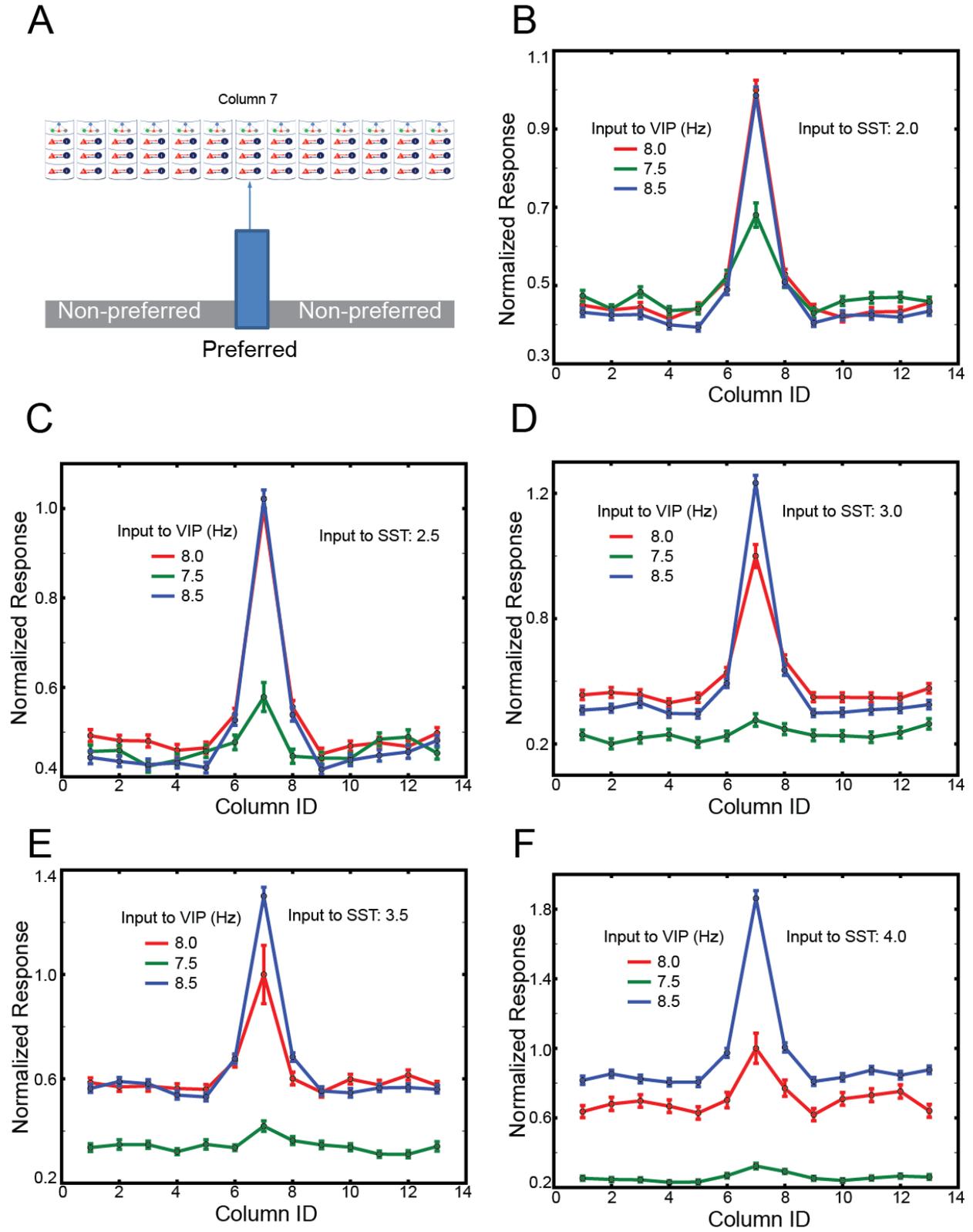

**S1 Fig.**

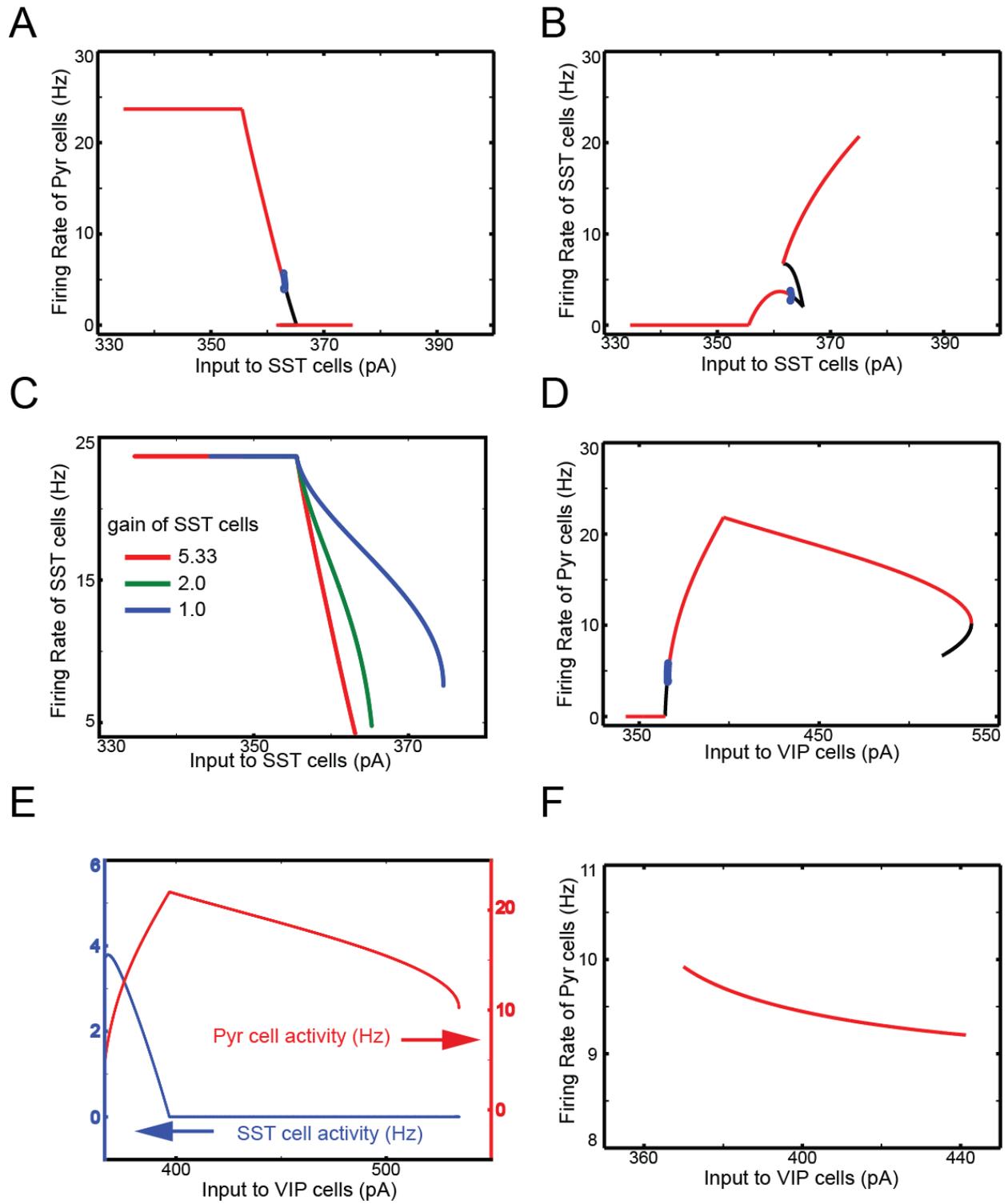

**S2 Fig.**

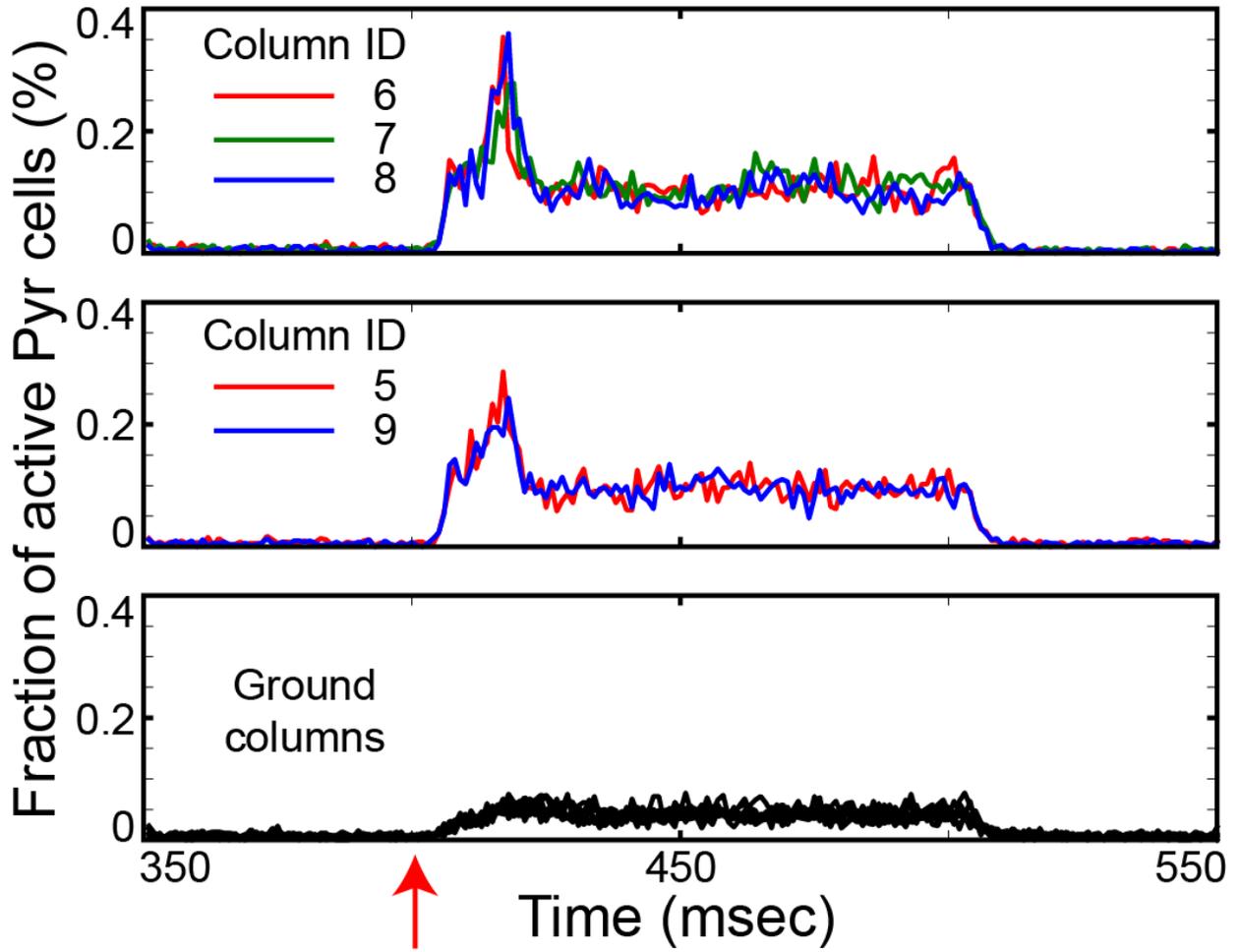

**S3 Fig.**

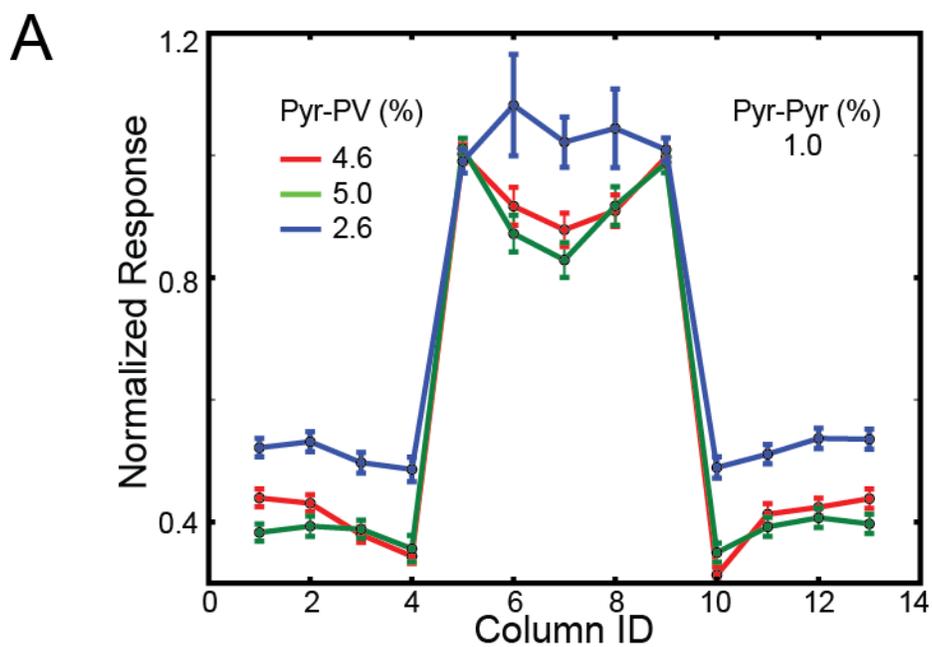

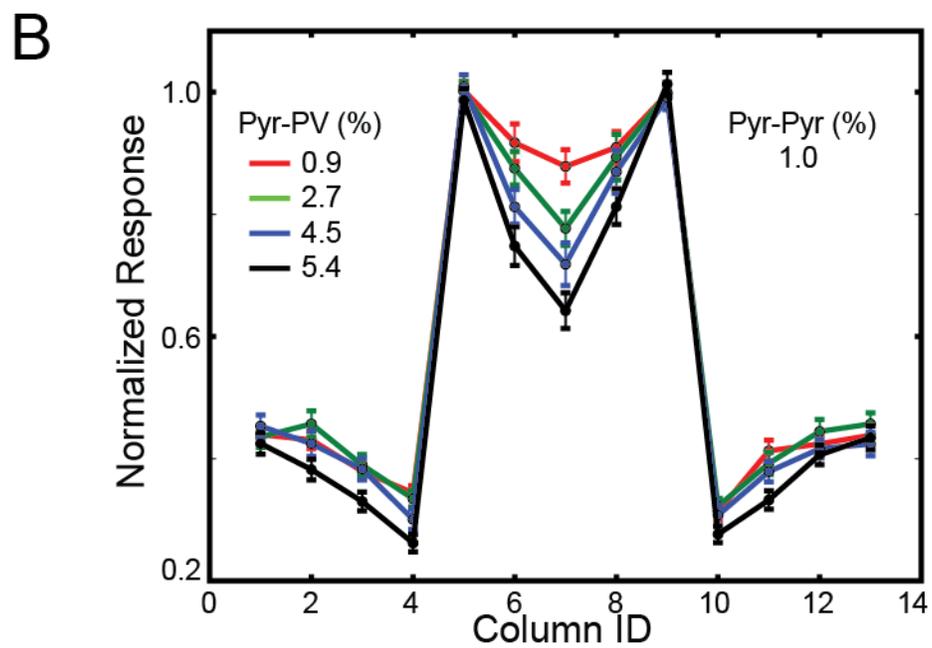

**S4 Fig.**

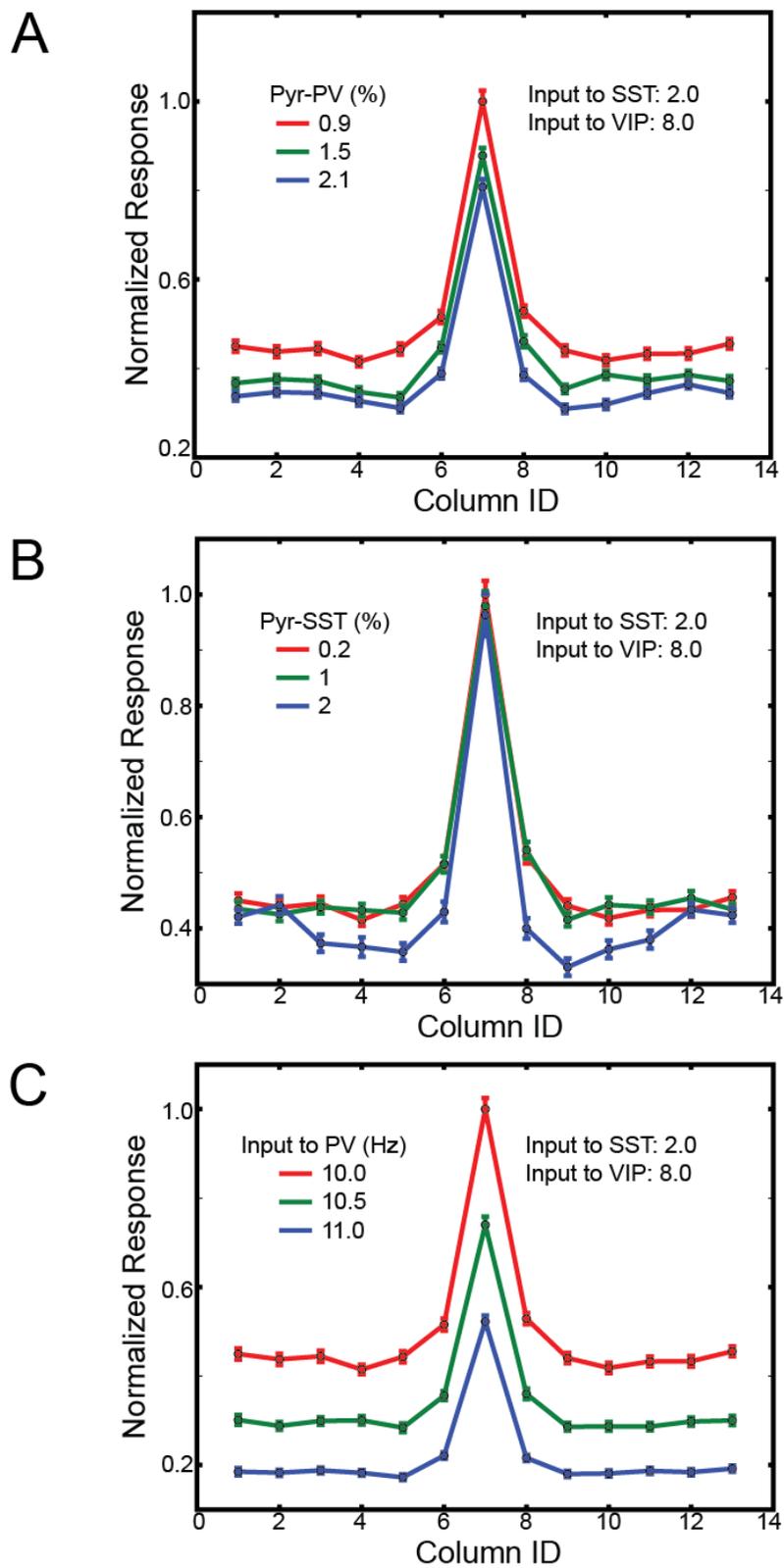